\documentclass[prl,superscriptaddress,a4paper,twocolumn,showpacs]{revtex4}
\usepackage{amsmath,amsfonts,graphicx,color}
\usepackage{amssymb}
\def\>{\rangle}
\def\<{\langle}
\def\u{\!\!\uparrow}
\def\U{\!\Uparrow}
\def\d{\!\!\downarrow}
\def\D{\!\Downarrow}
\def \be{\begin{equation}}
\def \ee{\end{equation}}
\def \beq{\begin{equation}}
\def \eeq{\end{equation}}
\def \bea{\begin{eqnarray}}
\def \eea{\end{eqnarray}}

\begin{document}

\title{A photonic cluster state machine gun}
\author{Netanel H. Lindner}%
\affiliation{Department of Physics, Technion---Israel Institute of Technology, 32000 Haifa,
Israel}

\author{Terry Rudolph}%
\affiliation{Optics Section, Blackett Laboratory, Imperial College
London, London SW7 2BZ, United Kingdom}%
\affiliation{Institute for Mathematical Sciences, Imperial College London, London SW7 2BW,
United Kingdom}%


\begin{abstract}
We present a method to convert certain single
photon sources into devices capable of emitting large strings of
photonic cluster state in a controlled and pulsed ``on demand''
manner. Such sources would greatly reduce the resources required to
achieve linear optical quantum computation. Standard spin errors,
such as dephasing, are shown to affect only 1 or 2 of the emitted
photons at a time. This allows for the use of standard fault
tolerance techniques, and shows that the photonic machine gun can be fired
for arbitrarily long times. Using realistic parameters for current
quantum dot sources, we conclude high entangled-photon emission
rates are achievable, with Pauli-error rates per photon of less than $0.2\%$.
For quantum dot sources the method has the added advantage of
alleviating the problematic issues of obtaining identical photons
from independent, non-identical quantum dots, and of exciton
dephasing.
\end{abstract}

\maketitle

The primary challenge facing optical quantum computation is
that of building suitable photon sources. The majority of effort
has been directed at single photon sources. Four single photons
can be used in an interferometer to produce a maximally
entangled Bell pair of photons \cite{pan}, and given a source of Bell pairs
it is in principle possible to fuse them \cite{Bro05} into larger
so-called \emph{cluster states} \cite{Rau01}. These somewhat
magical quantum states can be used for performing quantum
computation via the simple procedure of making individual
(single-qubit) measurements on the photons involved. Recently a
promising new approach has been  to produce Bell pairs directly
\cite{dudi,yosi} via a radiative cascade in quantum dots.
However, even an ideal such source would only reduce the overall
resources required for a full optical quantum computation by a small
factor.

We will show that with current technology it is
possible to manipulate certain single photon sources, in
particular quantum dots, so as to generate a continuous stream of
photons entangled in long strings of (various varieties of)
1-dimensional cluster states. Using these strings, cluster states
capable of running arbitrary quantum algorithms can be very
efficiently generated by fusion. We analyze all error mechanisms
and show that the error rates can be very low - close to fault
tolerant thresholds for quantum computing - even if the source is
operated for timescales much longer than the typical decoherence
times.

We begin with a highly idealized description of the proposal.
Consider a source with a degenerate spin $1/2$ ground state
manifold. The basis $|\u\>$, $|\d\>$ denotes the spin projection
along the $z$ axis. Furthermore, imagine that optical transitions
at frequency $\omega_0$ are possible \textit{only} to a doubly
degenerate excited state manifold. The excited states $|\U\>$,
$|\D\>$ have $J_z=\pm3/2\hbar$, thus only the (single photon)
transitions $|\u\>\!\leftrightarrow \! |\U\> $ and
$|\d\>\!\leftrightarrow \! |\D\>$ are allowed. Such transitions
are well known to occur, for example, in quantum dots (QDs) which
emit single photons via charged-exciton decay
\cite{bayer}. We only consider the emitted photons propagating along the $z$ axis. Therefore, if the initial state
of the source is $|\u\>$ ($|\d\>$), an excitation to the state
$|\U\>$ ($|\D\>$) followed by radiative decay, results in the
emission of a single right (left)-circularly polarized photon
$|R\>$ ($|L\>$) and leaves the source in the state $|\u\>$
($|\d\>$). Now, consider the initial state $|\u\>+|\d\>$, and a
coherent excitation pulse with a linear polarization along the $x$
direction. (The exciting pulse itself need not necessarily
propagate along the $z$ direction, which is useful for separation
of the coherent and emitted light). Such a pulse couples equally
to both transitions. Therefore, the processes described above
happen in superposition, and the emitted photon will be entangled
with the electron: the joint state of both systems would be the
Bell pair $|\u,R\>+|\d,L\>$. Repeating such a procedure would
produce GHZ-type entangled states, which are not useful for
quantum computing, and for which disentangling the photons from
the electron spin is difficult. Moreover, the GHZ state is highly
vulnerable to decoherence. By contrast, the cluster states suffer
none of these problems.

\begin{figure}
\includegraphics*[width=9.0cm, bb=2cm 6cm 22cm 17cm]{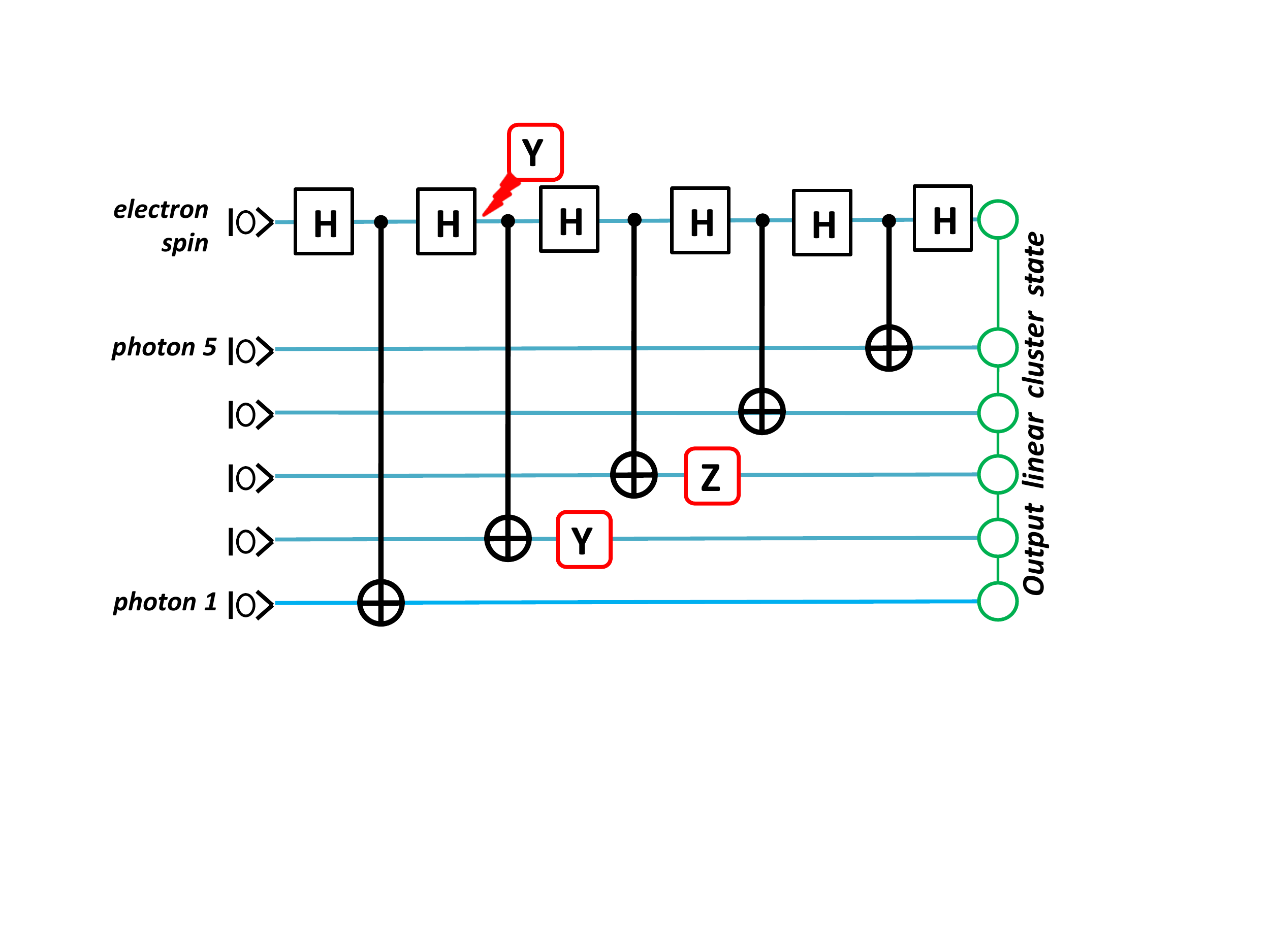}
\caption{(Color online) A  quantum circuit readily verified to output linear cluster state.
For mapping to the cluster state machine gun, the top qubit line
is the electron spin, the
Hadamard gates are replaced by single qubit unitaries $\exp(-i\pi
Y/4)$ (requiring the careful tracking of certain phases), and the physical process of creating a photon with
left/right circular polarization conditioned on the state of the
electron spin becomes the controlled not gate
which leaves the qubit (photon) in state $|0\>$ (i.e. $|R\>$) if
the electron spin is in state $|0\>$ (i.e. $|\u\>$), but otherwise
flips it. Crucially, as depicted, a Pauli $Y$ error on
the spin localizes; i.e it is equivalent to $Y$ and $Z$ errors on the next two photons
produced.}
 \label{circuit}
 \end{figure}

To see how to create cluster states, we now imagine that before
the second excitation of the system, when the state of the spin
and the first photon is $|\u\>|R_1\>+|\d\>|L_1\>$, the spin
undergoes a $\pi/2$-rotation about the y-axis. Under this
operation, described by $\exp(-iY\pi/4)$ the state evolves to $(|\u\>+|\d\>)|R_1\>+(-|\u\>+|\d\>)|L_1\>$.
A second pulse excitation,
accompanied by a second photon emission, will now result in the two photons
and the electron spin being in the state
$(|\u\>|R_2\>+|\d\>|L_2\>)|R_1\>+(-|\u\>|R_2\>+|\d\>|L_2\>)|L_1\>$. In terms of abstract (logical) qubit encodings we will take $|R\>\equiv|0\>,|L\>\equiv-|1\>$. It can be readily verified that rotating the spin with
another $\pi/2$ rotation, now leaves the spin and two photons in the  state:
$|000\>+|001\>+|010\>-|011\>+|100\>+|101\>-|110\>+|111\>,$
which is exactly the 3 qubit linear cluster state.
Repeating the process of excitation followed by $\pi/2$
rotation, will produce a third photon such that the electron and
three photons are in a 4-qubit linear cluster state. The
procedure can, in principle, be repeated indefinitely, producing a
continuous chain of photons in an entangled linear cluster state. Note that one advantage
of producing a cluster state is that the electron can be
readily disentangled from the string of entangled photons, for
example by making a computational ($|R\>,|L\>$) basis measurement
on the most recently created photon. In fact, since in general the
initial state of the spin will be mixed, such a detection of a photon in state $|R\>
(|L\>$) polarization can also be used to project the spin to the $|\u\>$ $(|\d\>)$
state, and initializes the cluster state (either outcome is ok). It can be readily verified that the whole idealized procedure just
described is equivalent to the qubit quantum circuit depicted in
Fig.~\ref{circuit}.

A general analysis of how cluster states are generated by evolution of atoms in cavities undergoing general pumping and decay can be found in \cite{schoen}, and interesting cavity QED proposals can be found in \cite{cavity}.
We will primarily focus on a specific
implementation of our proposal, namely photon emission from a
quantum dot, via the process of creation and subsequent decay of a
charged exciton (trion). In practise the expressions we derive,
such as the structure of the emitted photon wavepackets, can be
easily applied to any systems which obey similar selection rules,
and the imperfections we discuss are, for the most part, generic.
The importance of the selection rules arises as follows.
In semiconductor
quantum dots, the $J=3/2, J_z=\pm 1/2$ states are naturally split
off from the $J=3/2, J_z=\pm 3/2$ ones primarily due to confinement. They correspond to trions containing two electrons in the
singlet state and a
\textit{light} hole or \textit{heavy} hole respectively. We can consider only the \textit{heavy} trions and neglect
the mixing between them. In other systems, while the transitions
to $J_z=\pm 1/2$ may be energetically split off by an external
field, or may simply have different couplings, generically they
will still lead to imperfections equivalent to non-orthogonality
of the emitted photons. Moreover processes in other systems tend
to be slower, and temporally longer pulses may well also be
required because of nearby energy levels.  Although these problems
can be remedied somewhat by applying proper filtration protocols
to the output cluster state (at the expense of larger loss rates)
we focus on quantum dots for which the suppression is essentially
perfect, the processes are fast, and the energy levels well
separated.

Although other options exist, we will consider from now on the
situation where the $\pi/2$ rotations on the spin are performed by
placing the quantum dot in a constant magnetic field of strength
$B$ which is directed along the y-direction (i.e. in the plane of
the dot). The spin precession at frequency $\omega_B=g_e\mu
B/\hbar$  in the z-x plane therefore implements the desired
rotation every $T_{\rm cycle}=\pi/2\omega_B$. Suitably timed
strobing of the dot by the excitation pulse, followed by the rapid
exciton decay, will therefore enable the machine-gun-like
generation of 1d cluster state described above.

The potential imperfections to be considered are as follows: (i) The non-zero
lifetime of the trion $\tau_{\rm decay}$ means that the magnetic
field causes precession of the electrons during the emission
process. This leads to errors induced on the quantum circuit of
Fig~\ref{circuit}; however we
shall find that they can be understood as implementing an error
model on the final output cluster state which takes the form of
Pauli errors occurring with some independent probability on pairs
of (photonic) qubits. (ii) Interaction of the electron spin with
its environment results in a non-unitary evolution of the spin.
This evolution consists of two parts: decoherence (in which we
include both dephasing and spin flips) and spin relaxation.
Decoherence is characterized by a $T_2$ time.
Fortunately we will see that both these processes also lead only
to errors occurring independently on two (photonic) qubits at a
time.  Efficient cluster state quantum computation can proceed
even if every qubit has a finite (though small) probability of
undergoing some random error
\cite{clusterfaulttolerance}. This implies that the protocol's running time is not
limited by $T_2$, while the errors are amenable to standard
quantum error correction techniques for cluster states. Spin
relaxation is characterized by a $T_1$ time, and is a process
which projects the spin to the ground state. In semiconductor
quantum dots $T_1$ times are extremely long $T_1\gg
T_2\gg\tau_{\rm decay}$
\cite{experimentalT1}. Therefore, we shall not discuss the effects
of this process further here. We point out, however, that it can
be shown this process also leads to errors of a localized form,
and so in principle is no obstacle to the continuous operation of
the device even for times much longer than $T_1$. (iii) The last
source of error is related to the issue of ensuring the photons
are emitted into well-controlled spatial modes. In practise this
technological issue of mode matching (say by placing the dot in a
microcavity) results in some amount of photon loss error in the
final state. Significant progress on this issue is being made for
a variety of quantum dots
\cite{MICROCAVITYPAPERS,dirk}, although we emphasize that for our proposal strong coupling to the cavity is \emph{not} required. Fortunately photonic cluster state
computation can proceed even in the presence of very high (up to
50\%) loss \cite{Var08}, and we will not consider this source of
error further.

We now turn to detailed calculations of the error rate inflicted
by imperfections (i) and (ii) discussed above. We first calculate
the effect of a finite ratio of the trion decay time $\tau_{\rm
decay}$ time to the spin precession time.
We denote by $\rho_n(\tau+t_n)$ the state of the
system (the quantum dot and photons) at time $\tau$ after the
$n^{\rm th}$ excitation pulse, $t_n=nT_{\rm cycle}$. By
$\rho_{n}(t_n^-)$ we mean the state of the system
\textit{just before} the $n^{th}$ excitation pulse (we assume the
excitation is instantaneous). Following the excitation, the trion
state decays, emitting a photon and leaving an electron in the QD,
the spin of which then precesses in the magnetic field.
These lead
to an evolution of the quantum state described by the following
map (see \cite{epaps} for details):
\beq
\rho(t_n+\tau)=U^{\dag}(\tau)(G+F)^{\dag}\rho(t_n^-)(G+
F)U(\tau)
\label{rhon}
\eeq
The unitary operator $U=\exp(iY\omega_B\tau)\exp(i H_0
\tau)$ describes the precession of the electron spin and the free
propagation of the photons.
The generalized creation operators
$G^{\dag}=G^{\dag}_R|\u\>\<\;\u\!|+ G^{\dag}_L|\d\>\<\;\d\!|$,
$F^{\dag}= F^{\dag}_R|\d\>\<\;\u\!|- F^{\dag}_L|\u\>\<\;\d\!|$,
\label{photon error}
describe the excitation and decay process, adding a photon to the
state. The trion states decay exponentially with $\tau$, therefore
we have omitted them from Eq.~(\ref{rhon}) (which describes the
state of the system at times greater than the trion decay time,
\textit{i.e.} $\tau \gg \tau_{\rm decay}$). Note that the photons created in each cycle are well separated
from the ones created in the previous cycles (formally, this is
taken into account by the free propagation of the photons).

Equation (\ref{rhon}) describes a
circuit isomorphic to the one  in Fig.~\ref{circuit}. The
operator $G^{\dag}$  corresponds to a
correct application of a CNOT gate. This happens with an amplitude
$g(k)$, which depends on the photon's energy $k$:
$
\< k\varepsilon |G^{\dag}_\varepsilon|0\>\equiv g(k)=\frac{\sqrt{\Gamma/(2\pi)}(k-Z)}{(k-Z)^2-(g_e\mu
B)^2/4}. $
Here $|k\varepsilon\>=a^{\dag}_{k,\varepsilon}|0\>$ and
$\varepsilon=L,R$. The complex energy of the trion states is
denoted by $Z=\omega_0-i\Gamma/2$, where $\tau_{\rm
decay}=1/\Gamma$ is their lifetime. The operator $F^{\dag}$
corresponds to a CNOT gate followed by a $Y$ error on the spin
qubit. This errored gate is applied with amplitude $f(k)$, where $
\< k\varepsilon| F^{\dag}_\varepsilon|0\>\equiv f(k)=
\frac{i \sqrt{\Gamma/(2\pi)} g_e\mu B/2 }{(k-Z)^2-(g_e\mu B)^2/4}. $

\begin{figure}
\includegraphics*[width=9.0cm, bb=0.5cm 6cm 21cm 22cm]{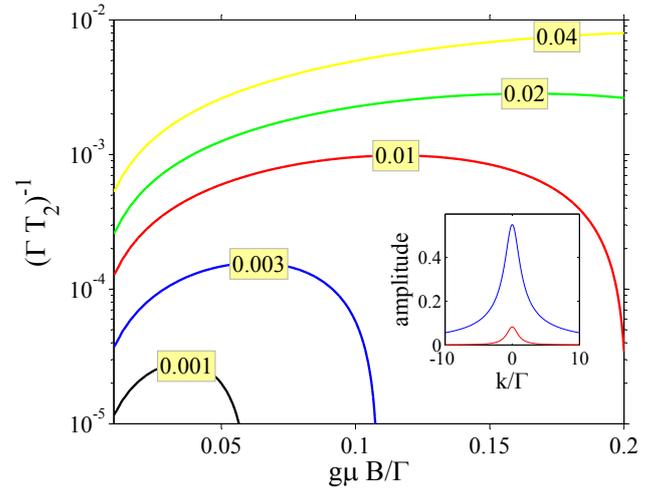}
\caption{(Color online) Contour plots showing the probability of Pauli error on any given photon, as a function of $g_e\mu B/\Gamma$ and
$(\Gamma T_2)^{-1}$. A stronger field causes faster precession
which increases a chance of the error during decay, but reduces
the standard dephasing error due to a finite $T_2$. The inset is a
plot of the good mode function $|g(k)|$ (blue) and error mode
function $|f(k)|$ (red) at $g_e\mu B/\Gamma=0.15$, from which it
can be deduced that spectral filtering can reduce the error rates
further.}
\vspace{-0.5cm}
\label{errordiagram}
\end{figure}

Let us for the moment treat the processes
described by $G^{\dag}$ and $F^{\dag}$ as incoherent with each
other. Then the resulting state is described by the circuit of
Fig.~\ref{circuit} with a probability
$
p_B=\|f\|^2=\frac{(g_e\mu B)^2}{2(g_e\mu B)^2+2\Gamma^2},
$
that each CNOT gate is followed by $Y$ error on the spin qubit. As
noted in Fig.~1, a state with a $Y$ error on the spin
after generation of the $n^{\rm th}$ photon (\textit{i.e}, after
the $n^{\rm th}$ CNOT), is equivalent to a state $Y$ and a $Z$
error on the $n^{\rm th}+1$
and $n^{\rm th}+2$ photons, with no error on the spin. 
 Note that
the error probability increases with magnetic field strength, because the
spin can precess more during the lifetime of the trion
\cite{sophia2005}. Therefore it is advantageous to consider
relatively low magnetic fields, for which $g_e\mu_B\ll \Gamma$.
Taking the coherence between $G^{\dag}$ and $F^{\dag}$ into account, it can be seen that a unitary correction $e^{iY\phi}$
with $\tan\phi =|\< g| f\>/\< g| g\>|$ yields a
further improvement of the error rate. We also point out that as $g(k)$ is more
localized around $\omega_0$ then $f(k)$ (inset of
Fig.~\ref{errordiagram}), selection of photons with energy
$|k-\omega_0|>\Delta$ would yield a lower error rate at the
expense of (heralded) loss.
%

The calculation above ignores the possibility of the
exciton dephasing \cite{excitondephasing} during the decay process. Pure dephasing, in
which both excited levels evolve the same (random) phase, will
have no affect on the entanglement in polarization with which we
are concerned. Cross dephasing (experimentally seen to be very small \cite{excitondephasing}) will lead to $Z$-errors on the
qubits, which also localize (See \cite{epaps} for a detailed discussion).

We now turn to the issue of decoherence of the spin as a result of
its interaction with the nuclei in the quantum dot.
Assuming Markovian dynamics (discussed further in \cite{epaps}), it is well known
\cite{NielsenChuang} that the resulting dephasing and spin flip
dynamics are equivalent to the action of random Pauli operations
$X,Y,Z$ with some probabilities  $p_x,p_y,p_z$.
The probabilities $p_x,p_z$ are suppressed due to the presence of
the magnetic field, while $p_y$ is characterized by
$T_2$, the dephasing time. This can readily be shown to give
$p_{y}=\frac{1}{2}(1-e^{-T_{\rm cycle}/T_2})$ as the probability
of a given spin error in 1 cycle. We already noted that a $Y$
error on the spin becomes a pauli error on the next 2 photons.
Similarly a $Z$ error at the end of the $n^{\rm th}$ cycle is
equivalent to a $Z$ error on the $n^{\rm th}+1$ photon, as can be
seen from Fig.~\ref{circuit} and the fact that the operator
$C_{\rm NOT}(Z_{\rm spin}\otimes I_{\rm photon}) $ and $(I_{\rm
spin}\otimes Z_{\rm photon})C_{\rm NOT}$ have similar actions on
the states $|00\>$ and $|10\>$. As $X=iZY$, an $X$ error again
affects only the next two photons to be generated.

In Fig.~\ref{errordiagram} we plot the total probability of error
on any given qubit, $1-(1-p_B)(1-p_y)$, as a function of the two
dimensionless parameters $g_e\mu B/\Gamma$ and $(\Gamma
T_2)^{-1}$. We include the aforementioned easily implemented unitary correction. Although free induction decay $T_2$ may be relatively
short in low magnetic fields, using a spin echo pulse at half way
along the cycle time can extend $T_2$ considerably,
and remove the dephasing caused by a wide distribution of nuclear
(Overhauser) magnetic fields (often termed inhomogeneous
broadening and characterized by a $T_2^*$ time). To estimate an
achievable error rate, we consider a decay time of
$1/\Gamma=100\;{\rm ps}$, and a dephasing time of $T_2=1\;\mu{\rm
s}$ with the addition of the spin echo pulses (a lower bound of
$3\;\mu{\rm s}$ have been measured in high magnetic fields
\cite{Greilich}). This gives $(\Gamma T_2)^{-1} = 10^{-4}$. From
Fig.~\ref{errordiagram}, one can deduce that a probability of
error less then $0.2\%$ can be achieved by applying a magnetic
field of $~15 {\rm mT}$ (we take $g_e=0.5$). We note that even
without the spin echo pulses, error rates of about $1\%$ are
achievable, which enables the production of considerable longer
and higher quality optical cluster states than those produced by
current methods.


So far we have considered pulse excitations that are timed to
coincide with (integer multiples of) $\pi/2$ rotations of the
spin. In fact it can be advantageous to sometimes wait for a full
$\pi$ rotation to occur. This has the effect of emitting
subsequent photons which are redundantly encoded \cite{Bro05}.
Fusing together such qubits gives a highly efficient method for
producing higher-dimensional cluster states which are universal for
quantum computing.
Photons which undergo fusion can be spectrally filtered
(via a suitable prism), such that if they fail to pass the filter
they can still be measured and removed from the cluster state. This filtering does not lead to an increase in loss error
rates, but simply decreases the overall success probability of the
fusion gates.

Current experiments produce photonic cluster states via spontaneous parametric downconversion \cite{Walther}, and would seem to be limited to producing 6 to 8 photon cluster states. Our proposal in principle can produce strings of thousands of photons; however initial experiments will be limited by collection efficiency. With the parameters above a simple analysis shows that we would need a collection+photodetection efficiency of about 18\% for a demonstration of on-demand 12-photon cluster states, where the full 12 qubits are expected to be detected about once every 10 seconds.
%

Finally, our proposal is suggestive of an efficient mechanism for entangling matter qubits \cite{cavity,sean}, and we feel this is a topic worthy of further investigation.

\begin{acknowledgments}
  \textit{Acknowledgments.}  TR acknowledges the support of the EPSRC, the QIP-IRC, and the US Army Research Office. We acknowledge numerous useful conversations with S. Economou and D. Gershoni.
\end{acknowledgments}
%
%
%

\newpage

\widetext

\section{Supplementary material for the photonic cluster state machine gun}

Note: In the published version this document accompanies the above paper as supplementary material in the form of an EPAPS archive - i.e it is the same material as can be found at reference [15] above. The references in this section are self contained and can be found at the end of this document.

\section{Calculation of emitted photon wavepackets}

In this section we compute the emitted photon wavepacket, taking into
account the fact that the electron spin is precessing during the emission.
This effect and the finite $T_{2}$ time of the electron, are the dominant
source of imperfect cluster state production in the machine gun. Any
notation not defined here is defined in the paper.

The Hamiltonian we consider is
\[
H=H_{0}+H_{int}
\]%
where $H_{0}=\omega _{0}P+H_{B}+H_{ EM}$. Here $P$ is a projector on the dot
excited state (trion) manifold and $H_{EM}$ is the free EM field Hamiltonian.

The Zeeman interaction for the electron is $H_{B}=g_{e}\mu_{B}\vec{S}%
_{e}\cdot \vec{B}$, where $\mu$ is the Bohr magneton, and $g_{e}$ is the
effective gyromagnetic ratio of the electron in the QD. For the low magnetic
fields we are dealing with, we can assume $g_h=0$ for the heavy hole [1]. We
abbreviate $\frac{1}{2}g_{e}\mu _{B}\rightarrow g_{e}$. We take $\vec{B}=B%
\hat{y}$. The interaction Hamiltonian of the QD with the photon field, $%
H_{int}$ is given in the rotating wave and dipole approximation by
\[
H_{int}=\sum_{k}V_{k}\left( \left\vert \uparrow \right\rangle \left\langle
\Uparrow \right\vert a_{R,k}^{\dag }+\left\vert \downarrow \right\rangle
\left\langle \Downarrow \right\vert a_{L,k}^{\dag }+h.c.\right)
\]%
where $V_k$ are the coupling constants which depend on the details of the
specific quantum dot in question, and the (heavy) trion states are denoted
by $|\Uparrow\rangle$, $|\Downarrow\rangle$ respectively.

We want to calculate matrix elements of the form:

\begin{equation}
\left\langle \uparrow ,R_{k}\right\vert U(t)\left\vert \Uparrow
\right\rangle =\frac{1}{2\pi i}\int dE\exp (-iEt)\left\langle \uparrow
R\right\vert \mathcal{G}(E)\left\vert \Uparrow \right\rangle  \label{contour}
\end{equation}%
Here \ $\left\vert R_{k}\right\rangle $ and $\left\vert L_{k}\right\rangle $%
\ \ are left and right circularly polarised single photon states created by $%
a_{R,k}^{\dag },a_{L,k}^{\dag },$ and

\[
\mathcal{G}(E)=\frac{1}{E-H}
\]%
is the resolvent of the hamiltonian.

Denote by $Q$ the projector on ground state manifold. Then the matrix block
inversion formula gives

\[
Q\mathcal{G}(E)P=\frac{Q}{E-QHQ}V\frac{P}{E-PH_{0}P-PR(E)P}
\]%
with $R(E)=V\frac{1}{E-QH_{0}Q}V.$

The operator $R(E)$ can only couple the state $\left\vert \Uparrow
\right\rangle $ (or $\left\vert \Downarrow \right\rangle )$ to itself
(recalling that the hole is assumed to not precess).

Therefore%
\[
\mathcal{G}^{h}(E)\equiv \left( \frac{P}{E-PH_{0}P+PR(E)P}\right)
_{ss^{\prime }}=\delta _{ss^{\prime }}\frac{1}{E-\omega _{0}-r(E)}
\]%
where $r(E)=\int dk\frac{\left\vert V_{k}\right\vert ^{2}}{E-k}$ , and $%
s,s^{\prime }=\left\vert \Uparrow \right\rangle ,\left\vert \Downarrow
\right\rangle $\ . When performing the contour integral ($1$), we can
replace $r(E)$ with the lamb shift $\Delta $ and the inverse decay rate $%
\Gamma $. In the following we shall take $V_{k}=\sqrt{\Gamma /2\pi }$ for
simplicity, and absorb $\Delta $ into the definition of $\omega _{0}.$ This
gives

\[
\mathcal{G}^{h}(E)=\delta _{ss^{\prime }}\frac{1}{E-\omega _{0}+i\Gamma /2}
\]

Next, we consider

\begin{eqnarray*}
\mathcal{G}^{e}(E) &\equiv &\left( \frac{Q}{E-QHQ}\right) _{\sigma ^{\prime
}\sigma }=\left(
\begin{array}{cc}
E-\hbar k & ig_{e}B \\
-ig_{e}B & E-\hbar k%
\end{array}%
\right) _{\sigma ^{\prime }\sigma }^{-1} \\
&=&\frac{1}{D^{e}(E)}\left(
\begin{array}{cc}
E-\hbar k & -ig_{e}B \\
ig_{e}B & E-\hbar k%
\end{array}%
\right) _{\sigma ^{\prime }\sigma }
\end{eqnarray*}%
where $D^{e}(E)=(E-\hbar k)^{2}-g_{e}^{2}B^{2}.$ We now have the final
result:%
\[
\left\langle \sigma ,\varepsilon _{k}\right\vert \mathcal{G}(E)\left\vert
s\right\rangle =V_{k}^{e}\mathcal{G}(E)_{\sigma \sigma ^{\prime
}=\varepsilon }^{h}\mathcal{G}(E)_{s^{\prime }=\varepsilon ,s}
\]

Writing the matrix elements explicitly:
\begin{eqnarray*}
\left\langle \uparrow ,R_{k}\right\vert \mathcal{G}(E)\left\vert \Uparrow
\right\rangle  &=&V_{k}\frac{(E-\hbar k)}{D^{e}(E)(E-\omega _{0}+i\Gamma /2)}
\\
\left\langle \downarrow ,R_{k}\right\vert \mathcal{G}(E)\left\vert \Uparrow
\right\rangle  &=&V_{k}\frac{ig_{e}B}{D^{e}(E)(E-\omega _{0}+i\Gamma /2)} \\
\left\langle \uparrow ,L_{k}\right\vert \mathcal{G}(E)\left\vert \Downarrow
\right\rangle  &=&V_{k}\frac{-ig_{e}B~}{D^{e}(E)(E-\omega _{0}+i\Gamma /2)}%
=-\left\langle \downarrow ,R_{k}\right\vert \mathcal{G}(E)\left\vert
\Uparrow \right\rangle  \\
\left\langle \downarrow ,L_{k}\right\vert \mathcal{G}(E)\left\vert
\Downarrow \right\rangle  &=&V_{k}\frac{(E-\hbar k)}{D^{e}(E)(E-\omega
_{0}+i\Gamma /2)}=\left\langle \uparrow ,R_{k}\right\vert \mathcal{G}%
(E)\left\vert \Uparrow \right\rangle
\end{eqnarray*}%
We now consider the contour integral. The two poles contributing in the
limit $t\rightarrow \infty $ are $E=k-g_{e}B$ and $E=k+g_{e}B.$ We denote $%
Z=\omega _{0}-i\Gamma /2$ and a standard complex integration gives%
\[
\left\langle \uparrow ,R_{k}\right\vert U(t)\left\vert \Uparrow
\right\rangle =\left[ \exp (-ig_{e}Bt)f_{1}+\exp (ig_{e}Bt)f_{2}\right]
\]%
where $f_{1}=\frac{\sqrt{\Gamma /2\pi }}{(k+g_{e}B-Z)},$ $f_{2}=\frac{\sqrt{%
\Gamma /2\pi }}{(k-g_{e}B-Z)}.$ Above and in the following we omit the phase
factors $e^{-ikt}$ . Using the notation
\begin{eqnarray*}
g(k) &=&f_{1}(k)+f_{2}(k)=\frac{\sqrt{\Gamma /2\pi }(k-Z)}{%
(k-Z)^{2}-(g_{e}B)^{2}} \\
f(k) &=&f_{1}(k)-f_{2}(k)=\frac{\sqrt{\Gamma /2\pi }g_{e}B}{%
(k-Z)^{2}-(g_{e}B)^{2}}
\end{eqnarray*}%
this simplifies to
\[
\left\langle \uparrow ,R_{k}\right\vert U(t)\left\vert \Uparrow
\right\rangle =\cos (g_{e}Bt)g(k)-i\sin (g_{e}B)f(k).
\]%
Similarly
\[
\left\langle \downarrow ,R_{k}\right\vert U(t)\left\vert \Uparrow
\right\rangle =i\exp (-ig_{e}Bt)f_{1}-i\exp (ig_{e}Bt)f_{2}
\]%
which yields
\begin{eqnarray*}
\left\langle \downarrow ,R_{k}\right\vert U(t)\left\vert \Uparrow
\right\rangle  &=&i\left[ -i\sin (g_{e}Bt)g(k)+\cos (g_{e}Bt)f(k)\right]  \\
&=&\sin (g_{e}Bt)g(k)+i\cos (g_{e}Bt)f(k)
\end{eqnarray*}%
An analogous calculation shows that the amplitudes for starting with the
down trion are%
\[
\left\langle \uparrow ,L_{k}\right\vert U(t)\left\vert \Downarrow
\right\rangle =-\left\langle \downarrow ,R_{k}\right\vert U(t)\left\vert
\Uparrow \right\rangle =-\sin (g_{e}Bt)g(k)-i\cos (g_{e}Bt)f(k)
\]%
and
\[
\left\langle \downarrow ,L_{k}\right\vert U(t)\left\vert \Downarrow
\right\rangle =\left\langle \uparrow ,R_{k}\right\vert U(t)\left\vert
\Uparrow \right\rangle =\left\langle \uparrow ,R_{k}\right\vert
U(t)\left\vert \Uparrow \right\rangle =\cos (g_{e}Bt)g(k)-i\sin (g_{e}B)f(k)
\]

  Starting from the ground state manifold, the excitation and
subsequent photon emission and spin precesion can be broken into "good" and
"bad" parts, by introducing the operators:

\begin{eqnarray*}
G^{\dag } &=&G_{R}^{\dag }\left\vert \uparrow \right\rangle \>\left\langle
\uparrow \right\vert \;\!\!\!+G_{L}^{\dag }\left\vert \downarrow
\right\rangle \left\langle \downarrow \right\vert , \\
F^{\dag } &=&F_{R}^{\dag }\left\vert \downarrow \right\rangle \left\langle
\uparrow \right\vert -F_{L}^{\dag }\left\vert \uparrow \right\rangle
\left\langle \downarrow \right\vert \>\;
\end{eqnarray*}

The operator $G^{\dag }$ corresponds to a correct application of a CNOT
gate. This happens with an amplitude $g(k)$, which depends on the photon's
energy $k$: $\langle k\varepsilon |G_{\varepsilon }^{\dag }|0\>\rangle
\equiv g(k),$ the operator $F^{\dag }$ corresponds to an errored gate which
is applied with amplitude $f(k)$, where $\langle k\varepsilon
|F_{\varepsilon }^{\dag }|0\rangle \equiv f(k).$

Starting from the state $|\uparrow \rangle ,$ we get

\[
\left\vert \psi \right\rangle =U_{\uparrow ,R_{k},\Uparrow }|\uparrow
,R_{k}\rangle +U_{\downarrow ,R_{k},\Uparrow }|\downarrow ,R_{k}\rangle
\]

which we break $\left\vert \psi \right\rangle $ into the good and bad
states: $\left\vert \psi \right\rangle =\left\vert \psi _{good}\right\rangle
+\left\vert \psi _{bad}\right\rangle $ as follows:

\begin{eqnarray*}
\left\langle \uparrow ,R_{k}\right\vert \psi _{good}\rangle &=&\cos \left(
g_{e}Bt\right) g(k) \\
\left\langle \downarrow ,R_{k}\right\vert \psi _{good}\rangle &=&\sin \left(
g_{e}Bt\right) g(k)
\end{eqnarray*}

and

\begin{eqnarray*}
\left\langle \uparrow ,R_{k}\right\vert \psi _{bad}\rangle &=&-i\sin \left(
g_{e}Bt\right) f(k) \\
\left\langle \downarrow ,R_{k}\right\vert \psi _{bad}\rangle &=&i\cos \left(
g_{e}Bt\right) f(k)
\end{eqnarray*}

Note that $\left\vert \psi _{bad}\right\rangle =\frac{f_{1}-f_{2}}{%
f_{1}+f_{2}}Y_{spin}\left\vert \psi _{good}\right\rangle $ where $%
Y_{spin}\left\vert \uparrow \right\rangle =i\left\vert \downarrow
\right\rangle $ and $Y_{spin}\left\vert \downarrow \right\rangle
=-i\left\vert \uparrow \right\rangle .$ The relation to the operators $G$
and $F$ is

\begin{eqnarray*}
\left\langle \uparrow ,R_{k}\right\vert \psi _{good}\rangle &=&\left\langle
\uparrow ,R_{k}\right\vert U(t)G\left\vert \uparrow \right\rangle \\
\left\langle \downarrow ,R_{k}\right\vert \psi _{good}\rangle
&=&\left\langle \downarrow ,R_{k}\right\vert U(t)G\left\vert \uparrow
\right\rangle
\end{eqnarray*}%
and

\begin{eqnarray*}
\left\langle \uparrow ,R_{k}\right\vert \psi _{bad}\rangle &=&\left\langle
\uparrow ,R_{k}\right\vert U(t)F\left\vert \uparrow \right\rangle \\
\left\langle \downarrow ,R_{k}\right\vert \psi _{bad}\rangle &=&\left\langle
\uparrow ,R_{k}\right\vert U(t)F\left\vert \uparrow \right\rangle
\end{eqnarray*}%
where $U(t)$ denotes the free propagation.

Similarily, if we start from the state $|\downarrow \rangle ,$ the resulting
state would be
\[
\left\vert \psi \right\rangle =U_{\uparrow ,L_{k},\Downarrow }|\uparrow
,L_{k}\rangle +U_{\downarrow ,L_{k},\Downarrow }|\downarrow ,L_{k}\rangle
\]%
Lets break $\left\vert \psi \right\rangle $ into good and bad states: $%
\left\vert \psi \right\rangle =\left\vert \psi _{good}\right\rangle
+\left\vert \psi _{bad}\right\rangle $ as follows:%
\begin{eqnarray*}
\left\langle \uparrow ,L_{k}\right\vert \psi _{good}\rangle &=&-\sin \left(
g_{e}Bt\right) g(k) \\
\left\langle \downarrow ,L_{k}\right\vert \psi _{good}\rangle &=&\cos \left(
g_{e}Bt\right) g(k)
\end{eqnarray*}%
and%
\begin{eqnarray*}
\left\langle \uparrow ,L_{k}\right\vert \psi _{bad}\rangle &=&-i\cos \left(
g_{e}Bt\right) f(k) \\
\left\langle \downarrow ,L_{k}\right\vert \psi _{bad}\rangle &=&-i\sin
\left( g_{e}Bt\right) f(k)
\end{eqnarray*}%
The relation to the operator $G$ and $F$ is similar to the one described
above.

Note that again $\left\vert \psi _{bad}\right\rangle =\frac{f_{1}-f_{2}}{%
f_{1}+f_{2}}Y_{spin}\left\vert \psi _{good}\right\rangle .$

If we start from the state $|\uparrow \rangle +|\downarrow \rangle ,$ the
resulting state would be

\[
\left\vert \psi \right\rangle =U_{y}(g_{e}Bt)(g(k)+f(k)Y_{spin})(|\uparrow
,R_{k}\rangle +|\downarrow ,L_{k}\rangle )
\]

where $U_{y}$ denotes spin rotation around the $y$ axis. The bad part of the
wavefunction is therefore equivalent to a Pauli $Y$ error on the ideal
cluster state - and as discussed in the paper this error on the spin can be
turned into errors on the either the two photons emitted the error occurs on
the spin, or as $Z$ errors on the photon which has already been emitted and
the first emitted photon. This latter possibility shows that this
\textquotedblleft quantum scapegoat\textquotedblright\ effect can act
backwards in time! Note that while the error is coherent, the coherence
between $\left\vert \psi _{good}\right\rangle $ and $\left\vert \psi
_{bad}\right\rangle $ can be removed by applying stabilizers randomly to the
photons. In fact the good and bad wavepackets have non-trivial overlap, and
in the next section we consider how our knowledge of this can allow us to
apply a simple unitary correction which increases the amplitude of ideal
cluster state produced.

\begin{figure}
\includegraphics*[width=7.0cm, bb=0cm 0cm 10cm 10cm]{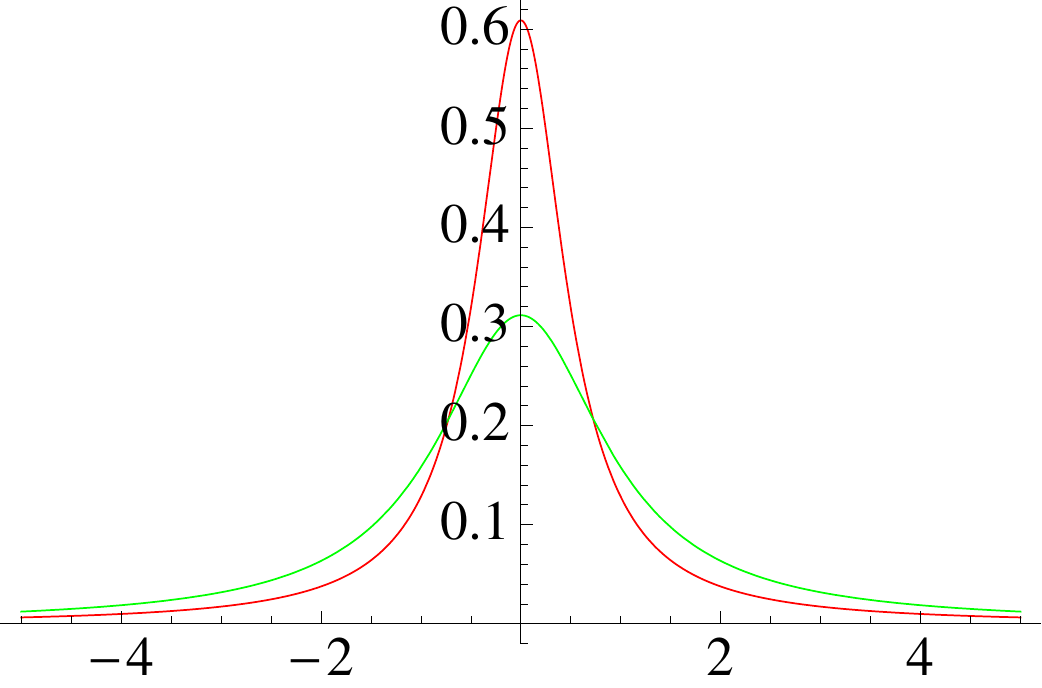}
\caption{Figure 1}
\end{figure}

\section{\textbf{Unitary correction}}

Considering the \textquotedblleft bad\textquotedblright\ amplitude $f(k)$ we
can clearly write it as

\[
f(k)=(f(k)-\alpha g(k))+\alpha g(k)
\]

with $\alpha =\langle g|f\rangle /\langle g|g\rangle .$ Note that first term
is orthogonal to $g(k).$Then

\[
\left\vert \psi \right\rangle =\frac{1}{\sqrt{N}}U_{y}(g_{e}Bt)[(1+\alpha
Y_{spin})g(k)\left( |\uparrow ,R_{k}\rangle +|\downarrow ,L_{k}\rangle
\right) +(f(k)-\alpha g(k))Y_{spin}\left( |\uparrow ,R_{k}\rangle
+|\downarrow ,L_{k}\rangle \right) ]
\]

(Note that $\alpha $ is purely imaginary.)

We can now make the correction $\exp (-i\phi Y_{spin})$ where $\tan \phi
=\alpha .$ Figure 2 of the paper is plotted assuming this simple form of
unitary correction has been carried out.

\begin{figure}
\includegraphics*[width=7.0cm, bb=0cm 0cm 10cm 10cm]{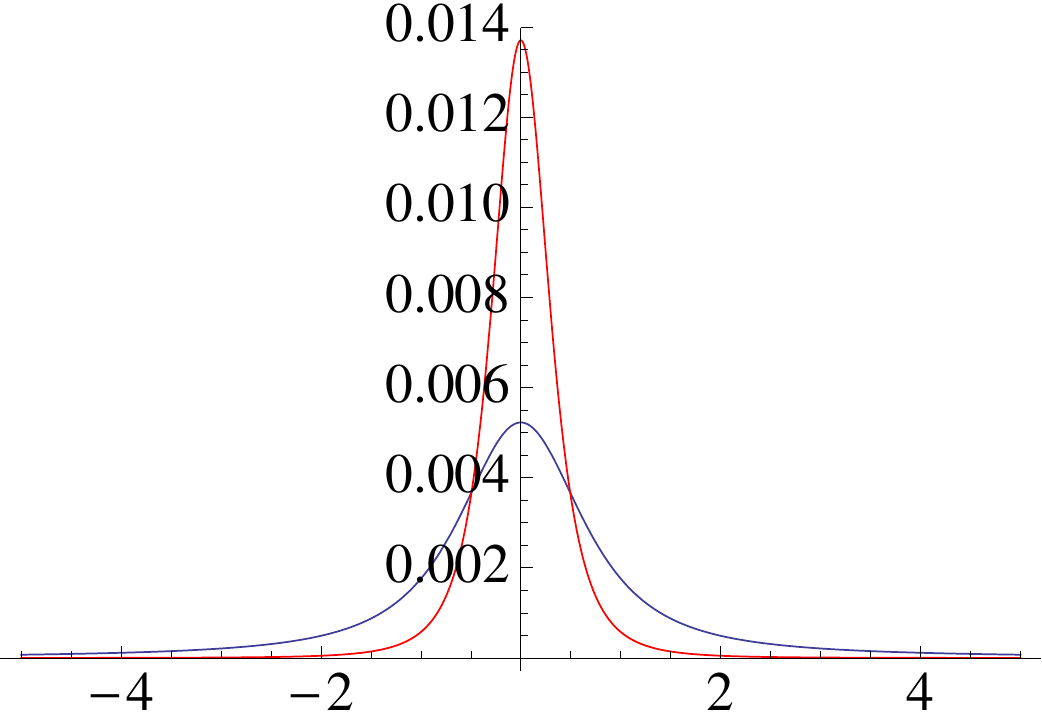}
\caption{Figure 2}
\end{figure}

\begin{figure}
\includegraphics*[width=7.0cm, bb=0cm 0cm 10cm 10cm]{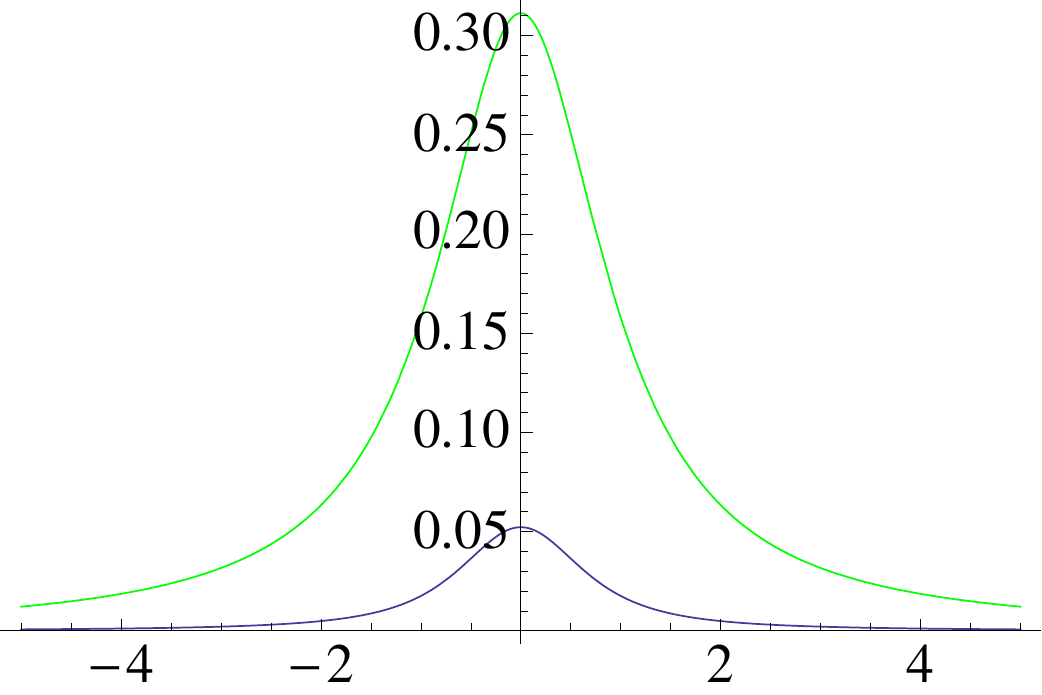}
\caption{Figure 3}
\end{figure}

\section{\textbf{Exciton Dephasing}}

In the paper we only briefly mentioned the spectral dephasing which will
occur while the system is excited. Our intuition contrasted with that of
others, namely we felt that this process would not affect the entanglement
of the state - in particular with respect to the polarization degrees of
freedom we are interested in - but would only lead to the emitted photon
wavepackets being in a mixture of different frequencies. This in turn would
only affect the (small fraction) of photons which have to go through fusion
gates, and such photons can be filtered before entering the gates in a way
which will only lead to a change in the success probability of the
(non-deterministic) gate. That is, such filtering need not even lead to a
loss error (as explained below). As such the only effect will be that we
need to use more photons - but the overhead is some constant factor.

Nevertheless, to be sure of how the device behaves and of the potential
magnitude of this effect, we turn now to the necessary calculations. We
begin with a description of the results and a heuristic discussion. There
are two excited state (exciton) energy levels to consider, which in the
paper we denote $\left\vert \Uparrow \right\rangle $, $\left\vert \Downarrow
\right\rangle $. These levels are degenerate. We consider both pure
dephasing and cross dephasing. In \emph{pure dephasing} of these levels,
their couplings to the environment are identical - in practise this means
that both states could evolve the same (random) phase. Note that this will
not affect the relative phase between the two states (such a relative phase
would ultimately become a relative phase between the two terms in the
entangled photonic cluster state). Therefore, it will have no bearing on the
quality of the entanglement in the polarization degrees of freedom. It will,
however, cause the emitted photons to have a broader range of frequencies
(broadened linewidth). In [2] such pure dephasing was measured, and from
their results one can infer that the new linewidth would be approximately
double the natural linewidth. However, we stress that this broadening is not
of particular importance to the operation of our protocol, for two reasons.
Firstly the majority of the emitted photons will never go through any
optical element where (for instance) they may be required to interfere with
other photons. Rather they will simply be measured directly, and this
polarization measurement will not care about their frequency. For the
photons that do need to undergo Type-II fusion gates (to cross link with
other qubit lines) the incoherence over the emitted frequencies can be
removed (if necessary) by suitable spectral filtering. This filtering can be
done in an \textquotedblleft active\textquotedblright\ manner with a
suitable prism wherein any photon which does not pass through the filter is
simply diverted into another path and measured in an appropriate basis - in
terms of the cluster state this simply removes the qubit. As such the only
effect of such imperfections is to change the probability of success of the
Type-II fusion gate. However efficient quantum computation is obviously
possible regardless how small this probability is, as long as it is finite.

Let us now show some calculations to bolster the above discussion. We treat
the pure exciton dephasing as a Markovian process and describe it using a
Lindblad operator $L=\sqrt{\gamma _{d}/2}(\left\vert \Uparrow \right\rangle
\langle \Uparrow \!|+|\!\Downarrow \rangle \langle \Downarrow \!|)$ where we
use the same notations as in the paper. The density matrix of the system
evolves according to a standard master equation. We compare the
probabilities described by this density matrix to the probabilities $%
|g(k)|^{2}$ corresponding to no error, and $|f(k)^{2}|$, corresponding to a
pauli error, which were calculated in the paper. In the presence of pure
dephasing, the probabilities corresponding to no pauli error are

\begin{eqnarray*}
|g_{\mathrm{dephase}}(k)|^{2} &=&\frac{\Gamma }{4\pi }\int_{0}^{\infty
}dt_{1}\int_{0}^{t_{1}}dt_{2}e^{(-\Gamma /2-\gamma _{d}/2)t_{1}}e^{(-\Gamma
/2+\gamma _{d}/2)t_{2}}\cos (bt_{1})\cos (bt_{2})e^{ik(t_{2}-t_{1})} \\
&+\frac{\Gamma }{4\pi }&\int_{0}^{\infty
}dt_{2}\int_{0}^{t_{2}}dt_{1}e^{(-\Gamma /2-\gamma _{d}/2)t_{2}}e^{(-\Gamma
/2+\gamma _{d}/2)t_{1}}\cos (bt_{1})\cos (bt_{2})e^{ik(t_{2}-t_{1})}
\end{eqnarray*}

In the above $b=g\mu B$ is the zeeman splitting, and we have shifted $%
k\rightarrow k-\omega _{0}$. The off-diagonal matrix elements corresponding
to pauli errors have the same form as Eq. but with $\sin $ functions
replacing the $\cos $ functions. Although the function $|g_{\mathrm{dephase}%
}(k)|^{2}$ is wider then $|g(k)|^{2}$, the total probability is equal to the
total probability without dephasing
\begin{equation}
\int dk|g_{\mathrm{dephase}}(k)|^{2}=\int dk|g(k)|^{2},
\end{equation}%
for any value $\gamma _{d}$.


In Figure 1 we plot the probabilities $|g_{\mathrm{dephase}}(k)|^{2}$ versus
$|g(k)|^{2}$. The parameters chosen for the plot are $b/\Gamma =0.15$ (as in
the inset of Fig. 2 of our manuscript) and $\gamma _{d}/\Gamma =1$ [1],
which of course enters only into $|g_{\mathrm{dephase}}(k)|^{2}$. Likewise,
Figure 2 plots $|f_{\mathrm{dephase}}(k)|^{2}$ vs. $|f(k)|^{2}$. In Figure
3, we compare $|g_{\mathrm{dephase}}(k)|^{2}$ and $|f_{\mathrm{dephase}%
}(k)|^{2}$ showing that spectral filtering is still possible also in the
presence of pure exciton dephasing.
%
%
%

A potentially more serious source of error would be \emph{cross dephasing}.
Experimentally such dephasing has not been observed, despite an attempt to
measure it in [2], where they could only lower bound the dephasing to be at
least 20 times smaller than the pure dephasing. Even if this cross dephasing
were exactly equal to the lower bound obtained in [2], the effect on the
emitted photons would ultimately lead to $Z$-errors on the qubits, which
also localize. This potential increase in the pauli error rate is negligible
compared to the other sources of pauli error we considered in the paper.

\section{\protect  \textbf{Note on Spin-Bath Dynamics}}

  To calculate the error probablilty on the photonic qubits we have
assumed Markovian dynamics for the interaction of the spin with its
environment. Previous studies [3-7] suggest a non-Markovian decay \ in time
of the coherence in the reduced density matrix of a spin coupled to a
nuclear spin bath. \ We note that for the short times we are interested in,
the Markovian assumption actually overestimates the single qubit error rate.
The resulting non markovian correlations of the errors on the different
qubits will be studied in a forthcoming publication. However, we note that
these correlations can be dealt with, see for example the discussion in [4].

 \bigskip

[1] J. G. Tischler\textit{\ et. al. ,} Phys. Rev. B \textbf{66}, 081310
(2002)

[2] A.J. Hudson et al, Phys. Rev. Lett. \textbf{99}, 266802 (2007)

[3] W. A. Coish\textit{\ et. al.}, Phys. Rev. B  \textbf{70}, 195340 (2004)

[4]  W. Yao \textit{et. al.}, Phys. Rev B \textbf{74}, 195301 (2006)

[5]   W. M. Witzel \textit{et. al.},Phys. Rev. Lett. \textbf{98}, 077601
(2007)

[6] F. H. L. Koppens \textit{et. al.}  Phys. Rev. Lett  \textbf{100}, 236802
(2008)

[7]  L. Cywisnski \textit{et. al. }Phys. Rev. Lett. \textbf{102}, 057601
(2009)

[8]\ B. Terhal and G. Burkard, Phys. Rev. A \textbf{71}, 012336 (2005)


\begin{thebibliography}{99}

\bibitem{pan} Q. Zhang, \textit{et. al}, Phys. Rev. A \textbf{77}, 062316 (2008)

\bibitem{Bro05} D. E. Browne and T. Rudolph
Phys. Rev. Lett. \textbf{95}, 010501 (2005).

\bibitem{Rau01} R. Raussendorf and H. J. Briegel,
\prl \textbf{86},  5188 (2001).

\bibitem{dudi} N. Akopian \textit{et. al}, Phys. Rev. Lett. \textbf{96}, 130501
(2006), M. Stevenson \textit{et. al.}, Nature
\textbf{439}, 179 (2006), A. Gilchrist \textit{et. al.}, Nature \textbf{445},
E4-E5.

\bibitem{yosi} E. Meirom \textit{et. al}., Phys. Rev. A 77, 062310 (2008),
J. E. Avron \textit{et. al}., Phys. Rev. Lett. \textbf{100} ,
120501 (2008).

\bibitem{bayer} M. Bayer \textit{et. al.} Phys. Rev. B \textbf{65}
195315 (2002).

\bibitem{schoen} C. Schoen et al, Phys. Rev. Lett. 95, 110503 (2005).

\bibitem{cavity} J. Metz \textit{et. al.}, Phys. Rev. A \textbf{76}, 052307 (2007);  C. Schoen \textit{et. al.}, Phys. Rev. A \textbf{75}, 032311 (2007); A. Beige \textit{et. al.}, J. Mod. Opt. 54, 397 (2007).

\bibitem{clusterfaulttolerance} R. Raussendorf \textit{et. al.}, Annals of Physics \textbf{321}, 2242 (2006); C. M. Dawson \textit{et. al.}, Phys. Rev. Lett. \textbf{96}, 020501 (2006); M. A. Nielsen and C. M. Dawson, eprint:quant-ph/0601066; Panos Aliferis and Debbie W. Leung, Phys. Rev. A \textbf{73}, 032308 (2006)

\bibitem{Var08} M. Varnava \textit{et. al},
Phys. Rev. Lett. \textbf{97}, 120501 (2006); ibid, Phys. Rev.
Lett.,\textbf{100}, 060502 (2008).

\bibitem{experimentalT1} M. Kroutvar \textit{et. al}., Nature \textbf{432},
81 (2004).

\bibitem{MICROCAVITYPAPERS} K. Hennessy \textit{et. al.}, Nature
\textbf{445}, 896 (2006); I. Fushman \textit{et. al.},
Phys. Stat. Sol. (c) \textbf{5}, 2808 (2008).



\bibitem{dirk} S. Strauf \textit{et. al.}, Nature Photonics \textbf{1}, 704 - 708
(2007); M. Rakher et al, Phys. Rev. Lett. \textbf{102} 097403 (2009).

\bibitem{epaps} See EPAPS document No. XXXX at http://www.aip.org/pubservs/epaps.html for detailed calculations of the emitted wavepackets and the effects of exciton dephasing.



\bibitem{sophia2005} S. E. Economou \textit{et. al.}, Phys. Rev. B \textbf{71}, 195327 (2005).

\bibitem{excitondephasing} A.J. Hudson \textit{et. al.}, Phys. Rev. Lett. \textbf{99}, 266802 (2007)




\bibitem{NielsenChuang} M. A. Nielsen and I. L. Chuang.
\textit{Quantum Computation and Quantum Information} (Cambridge
University Press, Cambridge, England, 2000).

\bibitem{Greilich} A. Greilich et al, Science \textbf{313}, 341 (2006).

\bibitem{Walther} P. Walther et al, Nature 434, 169-176 (2005); A.-N. Zhang et al, Phys. Rev. A 73, 022330 (2006).

\bibitem{sean} Sean D. Barrett and Pieter Kok, Phys. Rev. A \textbf{71} 060310(R) (2005)
\end{thebibliography}
\end{document}